\documentclass[preprint,number]{elsarticle}

\usepackage[T1]{fontenc}
\usepackage[latin1]{inputenc}
\usepackage[dvipdfm]{hyperref}
\usepackage{natbib}
\usepackage{amsmath}
\usepackage{nicefrac}
\usepackage{xspace}
\usepackage{wasysym}

\hypersetup{
pdftitle = {Analytical solutions of accreting black holes immersed in
  a Lambda-CDM model},
pdfauthor = {J. A. S Lima, Daniel C. Guariento, J. E. Horvath}
pdfkeywords = {black holes, cosmology}
}

\newcommand{\ud}{\ensuremath{\mathrm{d}}}

\title{Analytical solutions of accreting black holes immersed in a
  $\Lambda$CDM model}

\author[iag]{J. A. S. de Lima}
\ead{limajas@astro.iag.usp.br}

\author[if]{Daniel C. Guariento}
\ead{carrasco@fma.if.usp.br}

\author[iag]{J. E. Horvath}
\ead{foton@astro.iag.usp.br}

\address[iag]{Universidade de São Paulo -- Instituto de
  Astronomia, Geofísica e Ciências Atmosféricas\\
Rua do Matão, 1226, 05508-090 Cidade Universitária, São Paulo -- SP,
Brazil}

\address[if]{Universidade de São Paulo -- Instituto de Física\\
Rua do Matão, Travessa R, 187, 05508-090 Cidade Universitária, São
Paulo -- SP, Brazil}

\begin{document}

\begin{abstract}

The evolution of the mass of a black hole embedded in a universe
filled with dark energy and cold dark matter is calculated in a closed
form within a test fluid model in a Schwarzschild metric, taking into account
the cosmological evolution of both fluids. The result describes
exactly how accretion asymptotically switches from the
matter-dominated to the $\Lambda$-dominated regime. For early epochs,
the black hole mass increases due to dark matter accretion, and on
later epochs the increase in mass stops as dark energy accretion takes
over. Thus, the unphysical behaviour of previous analyses is improved
in this simple exact model.

\end{abstract}

\maketitle

\section{Introduction}

For the past decade, there has been overwhelming evidence of an
accelerated expansion of the universe. The simplest model proposed to
explain such a phenomenon is a fluid known as \emph{dark energy}, in 
which the parameters are best fit by Einstein's cosmological constant
\cite{wmap5-params}. Other models have also been proposed, many of
which have very interesting cosmological implications, such as
quartessence, which proposes a unified fluid with the characteristics
of both dark matter and dark energy, as well as more radical models
such as phantom (super-negative) energy.

In many cosmological models, the production of primordial black holes
is possible by a variety of mechanisms. It is still a matter of debate
whether these relics are present \cite{khlopov-2007} and play a role
as seeds to the formation of galaxies and galaxy clusters. There may
be a substantial evolution of the mass of black holes depending on
which epoch they were formed \cite{carr-2010,guariento-2007}
Therefore, it is of interest to have a detailed and accurate
description of such astrophysical objects along the evolution of the
universe.

In this work we study the evolution of black holes through accretion
of perfect fluids, by generalizing the approach to this problem started
by the works by Babichev \emph{et al} \cite{babichev-2004} and based
on the covariant conservation of the energy-momentum tensor in the
Schwarzschild metric. We consider the particular case of accretion of
dark matter and dark energy as described by the $\Lambda$CDM
scenario. We find a simple analytical model to describe the mass of
the black hole as a function of time which depends solely on the
initial values of the energy densities of dark matter and dark
energy. We also present specific descriptions of the model for
isolated physical species.

\section{Solution to the accretion equation}

\subsection{\texorpdfstring{The $\Lambda$CDM scenario}{The Lambda-CDM
    scenario}}

The most important constituents of our present universe are dark
energy (denoted with $\Lambda$ when modeled as a cosmological constant) and
cold dark matter (CDM). According to the WMAP5 data
\cite{wmap5-params}, these two species are responsible for 95\% of the
energy content in the universe, leaving the remaining 5\% for baryons,
neutrinos and radiation. Therefore, it is a sufficiently good
approximation to consider a universe containing only dark matter and
dark energy, provided we do not go too far into the past epochs, when
the contribution from the radiation field was relevant.

To describe the $\Lambda$CDM universe, we start with the local energy
conservation law of a cosmologically evolving fluid, which states

\begin{equation}\label{conservacao}
\dot{\rho} + 3 H  \left( \rho + p \right) = 0
\end{equation}

\noindent
so we have $\rho + p = - \nicefrac{\dot{\rho}}{3 H}$. From the Friedmann
equation $3 H^2 = 8 \pi G \rho$ we have

\begin{equation}\label{friedmann}
H = \left( \frac{8 \pi G}{3} \right)^{\nicefrac{1}{2}} \rho^{\nicefrac{1}{2}}.
\end{equation}

Thus, given the equation of state of the fluids, it is possible to
find $\rho (a)$ from \eqref{conservacao} and then insert the solution
on \eqref{friedmann} to find $a (t)$ \cite{weinberg-cosmology}.

In this work, we are interested in a universe filled with matter,
black holes and a cosmological constant, which represents quite
accurately the present state of our universe in the $\Lambda$CDM
model. Therefore, the energy density of the combined species is
$\rho_{\mathrm{t}} = \rho_{\text{DM}} + \rho_{\Lambda}$ (where black
holes are assumed never to be dynamically important), which, after
inserting their dependency on the scale factor, reads

\begin{equation}
\rho_{\mathrm{t}} = \rho^0_{\text{DM}} \left( \frac{a}{a_0}
  \right)^{-3} + \rho_{\Lambda}.
\end{equation}

The Friedmann equation \eqref{friedmann} for such a combination reads

\begin{equation}
\left( \frac{\dot{a}}{a} \right)^2 = H_0^2 \left[ \Omega_{\text{DM}}
  \left( \frac{a}{a_0} \right)^{-3} + \Omega_{\Lambda} \right]
\end{equation}

\noindent
and the solution is

\begin{equation}\label{lambdacdm}
a = a_0 \left[ \frac{\sinh \left( \frac{3}{2} H_0
    \sqrt{\Omega_{\Lambda}} t \right)}{\sqrt{
      \frac{\Omega_{\Lambda}}{\Omega_{\text{DM}}}}}
  \right]^{\nicefrac{2}{3}}.
\end{equation}

We use this result to compute the evolution of one test black hole
immersed in a universe containing these two species on the following
sections.

\subsection{The accretion equation}\label{sec-bab}

We start with the accretion equation written by Babichev \emph{et al.}
\cite{babichev-2004}, based on the conservation of the energy-momentum
tensor of a perfect non-self-gravitating fluid in the Schwarzschild
metric \cite{michel-1972}, and a mass variation term which can be
justified from geometrical properties of the energy-momentum tensor in
diagonal metrics \cite{guariento-2010}

\begin{equation}\label{mponto-bab}
\dot{m} = 4 \pi A m^2 \left[ \rho + p(\rho) \right]
\end{equation}

\noindent
where $\rho$ and $p$ are the energy density and pressure of the fluid
at infinity.

We neglect the effects of Hawking evaporation, which would otherwise
introduce a series of regime transitions from accretion to
evaporation \cite{guariento-2007}. Since evaporation only sets off
after a certain temperature threshold, this is equivalent to assuminng
that black holes are always colder ($T \propto \nicefrac{1}{m}$) than
the cosmic environment.

Equation \eqref{mponto-bab} may be rewritten through a change of
variables as

\begin{equation}\label{mpontoderho}
\frac{\ud m}{\ud \rho} \dot{\rho} = 4 \pi A m^2 \left( \rho + p
\right).
\end{equation}

After substituting the term $\left( \rho + p \right)$ using the
conservation equation \eqref{conservacao} on \eqref{mpontoderho}, we
find

\begin{equation}\label{mpontodeH}
\frac{\ud m}{\ud \rho} \dot{\rho} = - \frac{4 \pi A m^2 \dot{\rho}}{3
  H}
\end{equation}

\noindent
with a first integral

\begin{equation}\label{integrado}
-\frac{1}{m} = - \left( \frac{8 \pi}{3 G} \right)^{\nicefrac{1}{2}}
\rho^{\nicefrac{1}{2}} + C.
\end{equation}

To find the integration constant we set the initial value for the
black hole mass $m_{\mathrm{i}}$ at the instant with the initial fluid
density $\rho_{\mathrm{i}}$

\begin{equation}
C = \left( \frac{8 \pi}{3 G} \right)^{\nicefrac{1}{2}} A
\rho_{\mathrm{i}}^{\nicefrac{1}{2}} - \frac{1}{m_{\mathrm{i}}}.
\end{equation}

Inserting the value of the constant on equation \eqref{integrado} we
find the black hole mass as a function of the background density
\cite{sun-2008,martin-moruno-2008}

\begin{equation}\label{mderho}
m (\rho) = \frac{m_{\mathrm{i}}}{1 + m_{\mathrm{i}} \sqrt{\frac{8
      \pi}{3 G} A^2} \left( \rho^{\nicefrac{1}{2}} -
  \rho_{\mathrm{i}}^{\nicefrac{1}{2}} \right)}.
\end{equation}

We can also find the present value of the mass of a black hole under
these conditions by writing \eqref{mderho} in terms of
$m_0$. Inverting it, we are able to write the initial mass of a black
hole as a function of its present mass, provided we know the initial
value of the energy density of its main accreted component

\begin{equation}
m_i = \frac{m_0}{1 - m_0 \sqrt{\frac{8 \pi}{3 G} A^2} \left(
  \rho_0^{\nicefrac{1}{2}} - \rho_{\mathrm{i}}^{\nicefrac{1}{2}}
  \right)}.
\end{equation}

\section{Accretion of cosmological fluids}\label{fluidos}

\subsection{Cold dark matter (dust)}

For a universe filled only with non-relativistic matter,
$\rho_{\text{DM}} \propto a^{-3}$ one obtains the density from the
Einstein--de~Sitter model

\begin{equation}
\rho_{\text{DM}} = \rho^{\mathrm{i}}_{\text{DM}} \left(
\frac{t_{\mathrm{i}}}{t} \right)^2.
\end{equation}

\noindent
A black hole accreting matter in this scenario will evolve as

\begin{equation}\label{mdet-dm}
m_{\text{DM}} (t) = \frac{m_{\mathrm{i}}}{1 + m_{\mathrm{i}}
  \sqrt{\frac{8 \pi}{3 G} A^2 \rho^{\mathrm{i}}_{\text{DM}}} \left[
    \frac{t_{\mathrm{i}}}{t} - 1 \right]}.
\end{equation}

This behaviour may seem outrageous, as the mass diverges in a finite
time. However, it has been made clear
\cite{carr-1974,harada-2006,carr-2010} that such a result is an
artifact of the local test-fluid approximation which does not take
into account the back-reaction onto the black hole metric, the
consequence being the absence of an upper bound for the accretion of a
pressureless fluid. Therefore, it cannot be used at all for
arbitrarily high dark matter densities and time intervals.

On the low density limit, however, this approach may prove useful as
a mechanism to gauge the importance of different components to the
black hole mass growth. In fact, if one carries out these calculations
for realistic initial values of the dark matter density in the
matter-dominated era, it can be seen that the mass growth is rapidly
quenched by the cosmic evolution of the dark matter component and
never becomes important \cite{guariento-2007}.

\subsection{Dark energy only}

A universe filled with dark energy which behaves as a cosmological
constant will evolve according to the ``pure'' de~Sitter model, $a
\propto e^{H t}$. Such a component with an equation of state
$p_{\Lambda} = -\rho_{\Lambda}$ has a constant energy density, as can
be seen from equation \eqref{conservacao}. Therefore, according to
equation \eqref{mderho}, a black hole immersed in such a fluid
remains at a constant mass $m_{\Lambda}$.

Although the analysis from section \ref{sec-bab} cannot be applied to
the cosmological constant due to the $\dot{\rho} = 0$ term in
\eqref{mpontoderho}, the result is still consistent with the results
from Babichev \emph{et al.} \cite{babichev-2004} and the constant mass
parameter employed to derive the Schwarzschild--de~Sitter metric, as
expected.

\subsection{\texorpdfstring{Dark energy and cold dark matter
    ($\Lambda$CDM)}{Dark energy and cold dark matter (Lambda-CDM)}}

In a universe with a scale factor evolving as
\eqref{lambdacdm}, a black hole will accrete the total energy, so we
must use $\rho_{\mathrm{tot}} = \rho_{\text{DM}} + \rho_{\Lambda}$ in
equation \eqref{mderho}. Therefore, the matter density will evolve as

\begin{equation}\label{rhodmdet}
\rho_{\text{DM}} (t) = \rho_{\text{DM}}^{\mathrm{i}}
\frac{\Omega_{\Lambda}}{\Omega_{\text{DM}}} \frac{1}{\sinh^2 \left(
  \frac{3}{2} H_0 \sqrt{\Omega{\Lambda}} t \right)}
\end{equation}

\noindent
and the dark energy density will remain constant, so the mass will follow 
the following function of time, with $\rho_{\mathrm{DM}} (t)$ given by
\eqref{rhodmdet}

\begin{equation}\label{mdet-lcdm}
m_{\mathrm{tot}} (t) = \frac{m_{\mathrm{i}}}{1 + m_{\mathrm{i}}
  \sqrt{\frac{8 \pi}{3 G} A^2} \left\{ \left[\rho_{\Lambda} +
    \rho_{\text{DM}}^{\mathrm{i}} \right]^{\nicefrac{1}{2}} - \left[
    \rho_{\Lambda} + \rho_{\mathrm{DM}} (t)
    \right]^{\nicefrac{1}{2}} \right\}}.
\end{equation}

A close inspection to equation \eqref{mdet-lcdm} shows that, if one
takes an arbitrarily high initial value for the dark matter density,
one recovers the unphysical Einstein--de~Sitter evolution from
equation \eqref{mdet-dm}. Therefore, to avoid the same artifacts due
to neglecting the back-reaction, the initial time and asymptotic
densities must be set close enough to the dark energy-dominated era,
when the black hole is massive enough and the evolution of the scale
factor is fast enough to dilute the dark matter density. This
requirement is as important as neglecting the Hawking evaporation,
which also requires a sufficiently large mass.

\begin{figure}[!htp]
\centering
\includegraphics[width=.8\textwidth]{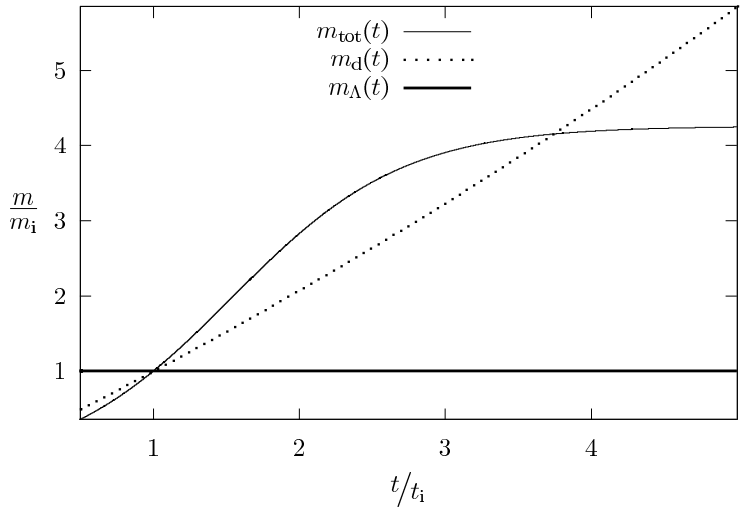}
\caption{Evolution of a black hole with an initial mass of
  $m_{\mathrm{i}} = 10^{-3} m_{\astrosun}$ in the three scenarios
  discussed on section \ref{fluidos}. Notice that accretion becomes
  negligible for $m_{\mathrm{tot}}(t)$ as dark energy
  dominates.}\label{thegraph}
\end{figure}

Figure \ref{thegraph} shows the evolution of the black hole masses
on the scenarios discussed above.

\section{Conclusion}

In this paper we have studied the evolution of black holes
which interact with a combination of dark matter and dark energy,
considering that in the vicinity of the black hole the metric is of
Schwarzschild type. This model represents a quasi-static accretion onto
a single black hole. We found the black hole mass as a function of time
by finding the energy densities of dark matter and dark energy as
solutions to the Friedmann equation and the local conservation law, in
which the black hole fluid is just a negligible fraction of the dark
matter content.

It is worth mentioning that the derivation of equation \eqref{mderho}
does not take into account the back-reaction of the accretion on the
surrounding metric, since it only considers a test fluid. Therefore,
some inaccuracies may arise when large $m_i$ and large variations of
it are computed. On the other hand, the considered black holes cannot 
be assumed too small either, since the accretion condition implies a 
flux of energy into them instead of the opposite physical situation 
represented by the Hawking evaporation
\cite{zeldovich-vol2,custodio-1998,barrow-1991}.
To achieve a more accurate description for a wider range of
conditions, one must consider the surrounding fluid also as
self-gravitating, and the full Einstein equations must be solved.

From the results presented on section \ref{fluidos}, we conclude that
black holes immersed in such a scenario feature an increase in mass,
but such an increase stops when dark energy becomes the most important
component. These results are consistent with the calculations 
performed by Babichev \emph{et al.} \cite{babichev-2004}, and they
also further generalize their model by taking into account the
evolution of the surrounding fluid with time.

\looseness=-1
Further developments of this approach must consider other types of
dark energy candidates, such as quintessence and quartessence, as well
as a more complete combination of fluids, which should provide a
clearer knowledge on the evolution of primordial black holes and
their influence on cosmology in general.

\section*{Acknowledgements}

The authors wish to thank C. E. Pellicer for some very helpful
discussions. JASL is supported by Conselho Nacional de Desenvolvimento
Científico e Tecnológico (CNPq-Brazil) and Fundação de Amparo à
Pesquisa do Estado de São Paulo (FAPESP), 2005/02809-5. DCG and JEH
are supported by CNPq-Brazil through grants and fellowships.

\bibliographystyle{elsarticle-num}
\bibliography{referencias}

\begin{thebibliography}{10}
\expandafter\ifx\csname url\endcsname\relax
  \def\url#1{\texttt{#1}}\fi
\expandafter\ifx\csname urlprefix\endcsname\relax\def\urlprefix{URL }\fi
\expandafter\ifx\csname href\endcsname\relax
  \def\href#1#2{#2} \def\path#1{#1}\fi

\bibitem{wmap5-params}
J.~Dunkley, E.~Komatsu, M.~R. Nolta, D.~N. Sperge, D.~Larson, G.~Hinshaw,
  L.~Page, C.~L. Bennett, B.~Gold, N.~Jarosik, J.~L. Weiland, M.~Halpern, R.~S.
  Hill, A.~Kogut, M.~Limon, S.~S. Meyer, G.~S. Tucker, Wollack, E.~L. Wright,
  Five-year {{Wilkinson Microwave Anisotropy Probe}} observations: Likelihoods
  and parameters from the {{WMAP}} data, The Astrophysical Journal Supplement
  Series 180~(2) (2009) 306--329.
\newblock \href {http://arxiv.org/abs/0803.0586} {\path{arXiv:0803.0586}},
  \href {http://dx.doi.org/10.1088/0067-0049/180/2/306}
  {\path{doi:10.1088/0067-0049/180/2/306}}.

\bibitem{khlopov-2007}
M.~Y. Khlopov, Primordial black holes, in: J.~A. de~Freitas~Pacheco (Ed.),
  Recent Advances on the Physics of Compact Objects and Gravitational Waves,
  Research Signpost, 2007.
\newblock \href {http://arxiv.org/abs/0801.0116} {\path{arXiv:0801.0116}}.

\bibitem{carr-2010}
B.~J. Carr, T.~Harada, H.~Maeda, Can a primordial black hole or wormhole grow
  as fast as the universe?, Classical and Quantum Gravity 27~(183101).
\newblock \href {http://arxiv.org/abs/1003.3324} {\path{arXiv:1003.3324}},
  \href {http://dx.doi.org/10.1088/0264-9381/27/18/183101}
  {\path{doi:10.1088/0264-9381/27/18/183101}}.

\bibitem{guariento-2007}
D.~C. Guariento, J.~E. Horvath, P.~S. Cust\'odio, J.~E. de~Freitas~Pacheco,
  Evolution of primordial black holes in a radiation and phantom energy
  environment, General Relativity and Gravitation 40~(8) (2008) 1593--1602.
\newblock \href {http://arxiv.org/abs/0711.3641} {\path{arXiv:0711.3641}},
  \href {http://dx.doi.org/10.1007/s10714-007-0562-8}
  {\path{doi:10.1007/s10714-007-0562-8}}.

\bibitem{babichev-2004}
E.~O. Babichev, V.~I. Dokuchaev, Y.~N. Eroshenko, Black hole mass decreasing
  due to phantom energy accretion, Physical Review Letters 93~(021102).
\newblock \href {http://arxiv.org/abs/gr-qc/0402089}
  {\path{arXiv:gr-qc/0402089}}.

\bibitem{weinberg-cosmology}
S.~Weinberg, Cosmology, Oxford University Press, 2008.

\bibitem{michel-1972}
F.~C. Michel, Accretion of matter by condensed objects, Astrophysics and Space
  Science 15~(1) (1972) 153--160.

\bibitem{guariento-2010}
D.~C. Guariento, J.~E. Horvath, On the problem of mass variation of black holes
  accreting cosmological fluids, (submitted for publication).

\bibitem{sun-2008}
C.-Y. Sun, Phantom energy accretion onto black holes in a cyclic universe,
  Physical Review D 78~(064060).
\newblock \href {http://dx.doi.org/10.1103/PhysRevD.78.064060}
  {\path{doi:10.1103/PhysRevD.78.064060}}.

\bibitem{martin-moruno-2008}
P.~M. Moruno, On the formalism of dark energy accretion onto black- and
  worm-holes, Physics Letters B 659 (2008) 40--44.

\bibitem{carr-1974}
B.~J. Carr, S.~W. Hawking, Black holes in the early universe, Monthly Notices
  of the Royal Astronomical Society 168 (1974) 399--415.

\bibitem{harada-2006}
T.~Harada, H.~Maeda, B.~J. Carr, Nonexistence of self-similar solutions
  containing a black hole in a universe with a stiff fluid or scalar field or
  quintessence, Physical Review D 74~(024024).

\bibitem{zeldovich-vol2}
Y.~B. Zel'dovich, I.~D. Novikov, Relativistic Astrophysics, Vol. 2: The
  Structure and Evolution of the Universe, The University of Chicago Press,
  Chicago, 1983.

\bibitem{custodio-1998}
P.~S. Cust\'odio, J.~E. Horvath, Evolution of a primordial black hole
  population, Physical Review D 58~(023504).

\bibitem{barrow-1991}
J.~D. Barrow, E.~J. Copeland, A.~R. Liddle, The evolution of black holes in an
  expanding universe, Monthly Notices of the Royal Astronomical Society 253
  (1991) 675--682.

\end{thebibliography}

\end{document}